\documentclass[sigconf]{acmart}

\copyrightyear{2021}
\acmYear{2021}
\setcopyright{acmcopyright}\acmConference[SIGIR '21]{Proceedings of the 44th International ACM SIGIR Conference on Research and Development in Information Retrieval}{July 11--15, 2021}{Virtual Event, Canada}
\acmBooktitle{Proceedings of the 44th International ACM SIGIR Conference on Research and Development in Information Retrieval (SIGIR '21), July 11--15, 2021, Virtual Event, Canada}
\acmPrice{15.00}
\acmDOI{10.1145/3404835.3462908}
\acmISBN{978-1-4503-8037-9/21/07}

\usepackage{multirow}
\usepackage{footmisc}
\usepackage{subcaption,siunitx,booktabs}
\usepackage{caption}
\usepackage{bbm}
\usepackage{balance}
\usepackage{enumitem}

\begin{document}
\fancyhead{}

\title{CauseRec: Counterfactual User Sequence Synthesis for Sequential Recommendation}

%\author{\vspace{-0.3cm}}

%\author[S. Zhang*, D. Yao*, Z. Zhao, T. Chua, F. Wu]{
%    Shengyu Zhang$^{1*}$, Dong Yao$^{1*}$, Zhou Zhao$^{1\dagger}$, Tat-seng Chua$^{2}$, Fei Wu$^{1\dagger}$
%}

\author{Shengyu Zhang$^{*}$}
\affiliation{
\institution{Zhejiang University, China}
\country{}
}\email{sy_zhang@zju.edu.cn}

\author{Dong Yao$^{*}$}
\affiliation{
\institution{Zhejiang University, China}
\country{}
}\email{yaodongai@zju.edu.cn}

\author{Zhou Zhao$^{\dagger}$}
\affiliation{
\institution{Zhejiang University, China}
\country{}
}\email{zhaozhou@zju.edu.cn}

\author{Tat-seng Chua}
\affiliation{
\institution{National University of Singapore}
\country{}
}\email{dcscts@nus.edu.sg}

\author{Fei Wu$^{\dagger}$}
\affiliation{
\institution{Zhejiang University, China}
\country{}
}\email{wufei@zju.edu.cn}

\renewcommand{\authors}{Shengyu Zhang, Dong Yao, Zhou Zhao, Tat-seng Chua, Fei Wu}

\renewcommand{\shortauthors}{Trovato and Tobin, et al.}
\newcommand{\etal}{\textit{et al}.}
\newcommand{\ie}{\textit{i}.\textit{e}.}
\newcommand{\eg}{\textit{e}.\textit{g}.}
\newcommand{\vpara}[1]{\vspace{0.05in}\noindent\textbf{#1 }}

\renewcommand{\thefootnote}{\fnsymbol{footnote}}


\begin{abstract}
Learning user representations based on historical behaviors lies at the core of modern recommender systems. Recent advances in sequential recommenders have convincingly demonstrated high capability in extracting effective user representations from the given behavior sequences. Despite significant progress, we argue that solely modeling the observational behaviors sequences may end up with a brittle and unstable system due to the noisy and sparse nature of user interactions logged. In this paper, we propose to learn accurate and robust user representations, which are required to be less sensitive to (attack on) noisy behaviors and trust more on the indispensable ones, by modeling counterfactual data distribution. Specifically, given an observed behavior sequence, the proposed CauseRec framework identifies dispensable and indispensable concepts at both the fine-grained item level and the abstract interest level. CauseRec conditionally samples user concept sequences from the counterfactual data distributions by replacing dispensable and indispensable concepts within the original concept sequence. With user representations obtained from the synthesized user sequences, CauseRec performs contrastive user representation learning by contrasting the counterfactual with the observational. We conduct extensive experiments on real-world public recommendation benchmarks and justify the effectiveness of CauseRec with multi-aspects model analysis. The results demonstrate that the proposed CauseRec outperforms state-of-the-art sequential recommenders by learning accurate and robust user representations \footnotetext[1]{These authors contributed equally to this work.}\footnotetext[2]{Corresponding Authors.}.

\end{abstract}





\maketitle

%\footnotetext{Code released at https://github.com/shengyuzhang/CauseRec}

\renewcommand{\thefootnote}{\arabic{footnote}}

\section{Introduction}

Due to the overwhelming data that people are facing on the Internet, personalized recommendation has become vital for retrieving information and discovering content. Accurately characterizing and representing users plays a vital role in a successful recommendation framework. Since users' historical interactions are sequentially dependent and by nature time-evolving, recent advances \cite{Manotumruksa_Yilmaz_2020,Ren_Liu_Li_Zhao_Wang_Ding_Wen_2020,Wang_Ding_Hong_Liu_Caverlee_2020,Wang_Fan_Xia_Zhao_Niu_Huang_2020,Ye_Wang_Chen_Wang_Qin_Yin_2020,Wang_Zhang_Rao_Qiu_Zhang_Lin_Zha_2020,Wang_Zhang_Ma_Liu_Ma_2020,Yuan_He_Karatzoglou_Zhang_2020,Ma_Ren_Lin_Chen_Ma_Rijke_2019,Zheng_Fan_Lu_Zhang_Yu_2019,Wang_Guo_Lan_Xu_Wan_Cheng_2015,Lu_Zhang_Huang_Wang_Yu_Zhao_Wu_2020} pay attention to sequential recommendation, which captures the current and recent preference by exploiting the sequentially modeled user-item interactions.

A sequential recommender aims to predict the next item a user might interact with based on the historical interactions. The challenging and open-ended nature of sequence modeling lends itself to a variety of diverse models. Traditional methods mainly exploit Markov chains \cite{Garcin_Dimitrakakis_Faltings_2013} and factorization machines \cite{Hidasi_Tikk_2016,Rendle_Freudenthaler_SchmidtThieme_2010} to capture lower-order sequential dependencies. Following these works, the higher-order Markov Chain and RNN (Recurrent Neural Network) \cite{He_Fang_Wang_McAuley_2016,Hidasi_Karatzoglou_Baltrunas_Tikk_2016,Wu_Ahmed_Beutel_Smola_Jing_2017} are proposed to model the complex high-order sequential dependencies. More recently, MIND is proposed to transform the historical interactions into multiple interest vectors using the capsule network \cite{Sabour_Frosst_Hinton_2017}. ComiRec \cite{Cen_Zhang_Zou_Zhou_Yang_Tang_2020} differs from MIND by leveraging the attention mechanism and introducing a factor to control the balance of recommendation accuracy and diversity.

Despite significant progress made with these frameworks, there are some challenges demanding further explorations. A vital challenge comes from the noisy nature of implicit feedback. Due to the ubiquitous distractions that may affect the users' first impressions (such as caption bias \cite{Lu_Zhang_Ma_2018}, position bias \cite{Jagerman_Oosterhuis_Rijke_2019}, and sales promotions), there are inconsistencies between users’ interest and their clicking behaviors, known as the natural noise \cite{OMahony_Hurley_Silvestre_2006}. Another challenge relates to the deficiency of existing methods in confronting data sparsity problem in recommender systems where users in general only interact with a limited number of items compared with the item gallery which can easily reach 100 million in large live systems. Therefore, solely modeling the observational behavior sequences that can be both sparse and noisy may end up with a brittle system that is less satisfactory. To this end, learning \textbf{accurate} and \textbf{robust} users' user representations is essential for recommender systems.

%--------------------------------fig-------------------------
\begin{figure}[!t] \begin{center}
    \includegraphics[width=0.9\columnwidth]{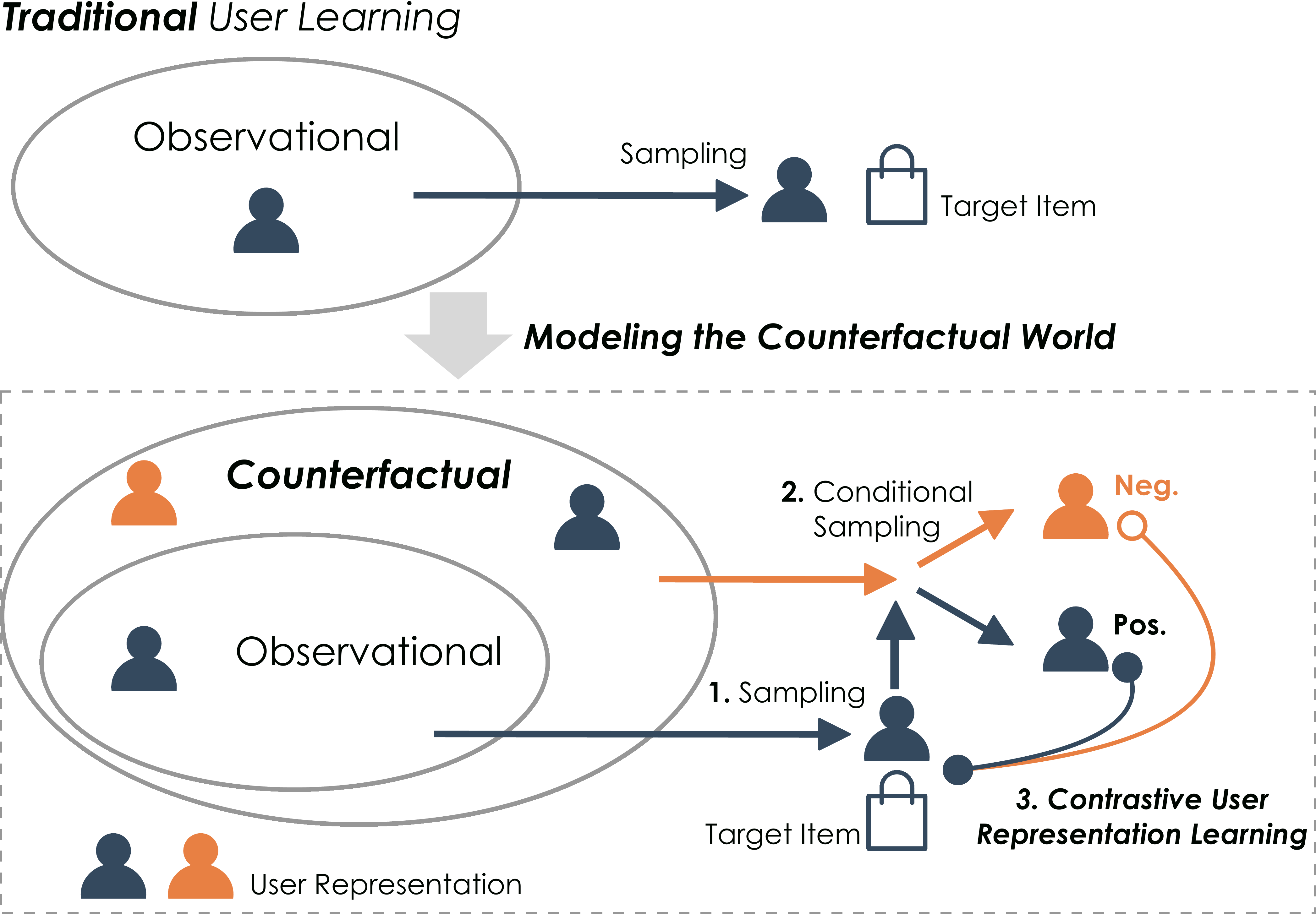}
    \caption{
    An illustration of the proposed contrastive user representation learning by modeling the counterfactual world (below), compared with most traditional approaches that solely model the observational user sequences (above).
    	}
\vspace{-0.4cm}
\label{fig:firstpage}
\end{center} \end{figure}
%--------------------------------fig end--------------------	

\iffalse Para4

To this end, we propose the COunterfactual INterest Network for Sequential Recommendation, abbreviated as CauseRec. 
	1. CauseRec identifies critical/inessential samples (换词)
	2. CauseRec build the counterfactual user representations (Deconstructive/Robust)
	3. CauseRec leans robust interest learners by contrastive learning

\fi

In this paper, we propose \textit{\textbf{C}}ounterf\textit{\textbf{a}}ctual \textit{\textbf{U}}ser \textit{\textbf{Se}}quence Synthesis for Sequential Recommendation, abbreviated as \textbf{CauseRec}. The essence of CauseRec in confronting the data sparsity problem is to model the counterfactual data distribution rather than the observational sparse data distribution where the latter can be a subset of the former one, as shown in Figure \ref{fig:firstpage}. We mainly aim to answer the counterfactual question, "what the user representation would be if we intervene on the observed behavior sequence?". Specifically, given the observed behavior sequence, we identify indispensable/dispensable concepts at both the fine-grained item level and the abstract interest level. A concept indicates a certain aspect of user interest/preference. We perform counterfactual transformations on both the item-level and the interest-level user concept sequences. We obtain 
counterfactually positive user representation by modifying dispensable concepts, and counterfactually negative user representation by replacing indispensable concepts. To learn \textbf{accurate} and \textbf{robust} user representations, we propose to conduct contrastive learning between: 1) the observational and the counterfactual user representations; and 2) the user representations and the target items. Contrast with such out-of-distribution hard negatives potentially makes the learned representations \textit{\textbf{robust}} since they are less sensitive to dispensable/noisy concepts. Contrast with such out-of-distribution positives potentially makes the learned representations \textit{\textbf{accurate}} since they will trust more on the indispensable concepts that are better representing user's interest.

We conduct in-depth experiments to validate the effectiveness of the proposed CauseRec architectures on various public recommendation datasets. With a naive deep candidate generation (or matching) architecture as the baseline method, CauseRec outperforms SOTA sequential recommenders for deep candidate generation. We conduct comprehensive model analysis to uncover how different building blocks and hyper-parameters affect the performance of CauseRec. Case studies further demonstrate that CauseRec can help learn accurate user representations. To summarize, this paper makes the following key contributions:
\begin{itemize}
\item We propose to model the counterfactual data distribution (besides the observational data distribution) to confront the data sparsity problem for recommendation.
\item We devise the CauseRec framework which learns \textit{accurate} and \textit{robust} user representations with counterfactual transformations on both fine-grained item-level and abstract interest-level, and with various contrastive objectives.
\item We conduct extensive experiments and show that with a naive deep candidate generation architecture as the baseline, CauseRec can outperform SOTA sequential recommenders.
\end{itemize}

\section{Related Works}

\subsection{Sequential Recommendation}

Sequential recommendation can be traced back to leveraging Markov-chain \cite{Garcin_Dimitrakakis_Faltings_2013,Feng_Li_Zeng_Cong_Chee_Yuan_2015} and factorization machines \cite{Rendle_Freudenthaler_SchmidtThieme_2010,Hidasi_Tikk_2016}. To capture long-term and multi-level cascading dependencies, deep learning based techniques (\eg, RNNs \cite{Hidasi_Karatzoglou_Baltrunas_Tikk_2016,Wu_Ahmed_Beutel_Smola_Jing_2017,Quadrana_Karatzoglou_Hidasi_Cremonesi_2017,feng2019temporal} and CNNs \cite{Tang_Wang_2018,Yuan_Karatzoglou_Arapakis_Jose_He_2019}) are incorporated into sequential modeling. DNNs are known to have enticing representation capability and have the natural strength to capture comprehensive relations \cite{zhang2021doc} over different entities (\eg, items, users, interactions). Recently, there are works that explore advanced techniques, \eg, memory networks \cite{Sukhbaatar_Szlam_Weston_Fergus_2015}, attention mechanisms \cite{Vaswani_Shazeer_Parmar_Uszkoreit_Jones_Gomez_Kaiser_Polosukhin_2017,Zhang_Tan_Zhao_Yu_Kuang_Jiang_Zhou_Yang_Wu_2020}, and graph neural networks \cite{Kipf_Welling_2017,feng2019graph,jin2021hierarchical,zhang2020does,lin2021bertgcn} for sequential recommendation \cite{Wang_Hu_Cao_Huang_Lian_Liu_2018,Ying_Zhuang_Zhang_Liu_Xu_Xie_Xiong_Wu_2018,Chen_Xu_Zhang_Tang_Cao_Qin_Zha_2018,Huang_Zhao_Dou_Wen_Chang_2018,Wu_Tang_Zhu_Wang_Xie_Tan_2019,Kang_McAuley_2018,Sun_Liu_Wu_Pei_Lin_Ou_Jiang_2019}. Typically, MIND \cite{Li_Liu_Wu_Xu_Zhao_Huang_Kang_Chen_Li_Lee_2019} adopts the dynamic routing mechanism to aggregate users' behaviors into multiple interest vectors. ComiRec \cite{Cen_Zhang_Zou_Zhou_Yang_Tang_2020} differs from MIND by leveraging the attention mechanism for user representations and proposes a factor for the trade-off between recommendation diversity and accuracy. Different from the above works that solely model the observational user sequences, we take a step further to model the counterfactual data distributions. By contrasting the user representations of the observation with the counterfactual, we aim to learn user encoders that can better confront out-of-distribution user sequences and learn accurate and robust user representations.

\subsection{Contrastive Learning for Recommendation}

A growing number of attempts have been made to exploit the complementary power of self-supervised learning (\eg, contrastive learning) and deep learning, with domains varying from computer vision \cite{He_Fan_Wu_Xie_Girshick_2020,Wu_Xiong_Yu_Lin_2018,feng2020shot}, natural language generation \cite{Clark_Luong_Le_Manning_2020,zhang2021drop}, to graph embedding \cite{Qiu_Chen_Dong_Zhang_Yang_Ding_Wang_Tang_2020}. However, how to consolidate the merits of contrastive learning into recommendation remains largely unexplored in the literature. Recently, Sun \etal \cite{Li_Tao_Zhang_Yu_Wang_2017} adopt noise contrastive estimation \cite{Gutmann_Hyvrinen_2010} to transfer the knowledge from a large natural language corpus to recommendation-specific content that is sparse on long-tail publishers and thus learning effective word representations. CLRec \cite{Zhou_Ma_Zhang_Zhou_Yang_2020} bridges the theoretical gap between contrastive learning objective and traditional recommendation objective, \eg, sampled softmax loss, as well as more advanced inverse propensity weighted (IPW) loss. They show that directly performing contrastive learning can help to reduce exposure bias. CP4Rec \cite{Xie_Sun_Liu_Gao_Ding_Cui_2020} and S$^3$-Rec \cite{Zhou_Wang_Zhao_Zhu_Wang_Zhang_Wang_Wen_2020} integrates Bert structure and contrastive learning objective for user pretraining, which require a fine-tuning stage. Compared with these works, we design model-agnostic and non-intrusive frameworks that help any baseline model learn more effective user representations in an end-to-end manner. Such representations are more accurate and robust by contrasting the original user representation with counterfactually positive samples and counterfactually negatives samples.

\subsection{Counterfactual for Recommendation}

Causality and counterfactual reasoning have attracted great attentions in various domains \cite{feng2021causalgcn,Zhang_Jiang_Wang_Kuang_Zhao_Zhu_Yu_Yang_Wu_2020,zhang2020counterfactual}. Previous counterfactual frameworks in recommendation focus on debiasing the learning-to-rank problems. A rigorous counterfactual learning framework, \ie, PropDCG \cite{Agarwal_Takatsu_Zaitsev_Joachims_2019}, is proposed to overcome the distorting effect of presentation bias. The position bias and the clickbait issue are investigated in \cite{Agarwal_Zaitsev_Wang_Li_Najork_Joachims_2019,Wang_Golbandi_Bendersky_Metzler_Najork_2018} and \cite{Wang_Feng_He_Zhang_Chua_2020}, respectively. The Inverse Propensity Score \cite{Schnabel_Swaminathan_Singh_Chandak_Joachims_2016,Yang_Cui_Xuan_Wang_Belongie_Estrin_2018} method obtains unbiased estimation by sample re-weighting based on the likelihood of being logged. Another line of works encapsulates the uniform data into recommendation by learning imputation models \cite{Yuan_Hsia_Yang_Zhu_Chang_Dong_Lin_2019}, computing propensity \cite{Schnabel_Swaminathan_Singh_Chandak_Joachims_2016}, using knowledge distillation \cite{Liu_Cheng_Dong_He_Pan_Ming_2020,feng2020kd3a}, and directly modeling the uniform data \cite{Bonner_Vasile_2018,Kallus_Puli_Shalit_2018,Liu_Rogers_Shiau_Kislyuk_Ma_Zhong_Liu_Jing_2017,Rosenfeld_Mansour_Yom_Tov_2017}. Different from these works, we focus on denoising user representation learning and considers the retrospect question, \ie, 'what the user representation would be if we intervene on the observed behavior sequence?'. Technically, we propose several counterfactual transformations based on the identification of indispensable/dispensable concepts and devise several contrasting objectives for learning accurate and robust user representations.

\section{Methods}

\subsection{Problem Formulation} \label{sec:problemformulation}

In the view of sequential recommendation, datasets can be formulated as $\mathcal{D}=\{(x_{u,t}, y_{u,t}) \}_{u=1,2,\ldots,N, t=1,2,\ldots,T_u}$, where $x_{u,t}=\{y_{u,1:(t-1)}\}$
denotes a user's historical behaviors prior to the $t$th behavior $y_{u,t}$ and arranged in a chronological order, and $T_u$ denotes the number of behaviors for the user $u$.
The goal of sequential recommendation is to predict the next item $y_{u,t}$ given the historical behaviors $x_{u,t}$, which can be formulated as modeling the probability of all possible items:
\begin{align}
	p\left( y_{u,t}=y | x_{u,t} \right),
\end{align}
We will drop the sub-scripts occasionally and write $(x, y)$ in place of $(x_{u,t}, y_{u,t})$ for simplicity.
Let $\mathcal{X}$ denote the set of all possible click sequences, i.e.\ $x\in \mathcal{X}$ and
each $y\in\mathcal{Y}$ represent a clicked item, while $\mathcal{Y}$ is the set of all possible items.

Since the number of items $|\mathcal{Y}|$ can easily reach 100 million, industrial recommender systems consist typically of two phases, \ie, the matching phase and the ranking phase, due to concerns on system latency. The matching (also called deep candidate generation) phase focuses on retrieving Top N candidates for each user, while the ranking phase further sorts the N candidates by typically considering more fine-grained user/item features and incorporating complex modeling architectures. In this paper, we mainly conduct experiments in the matching stage (\eg, comparing with SOTA matching models). 
%We note that the proposed CauseRec is designed as model-agnostic and can be incorporated into ranking models. 

\subsection{A Naive Matching Baseline} \label{sec:base}

The paradigm of a matching model includes a user encoder $f_{\theta}(x)\in \mathbb{R}^d$, which takes the user's historical behavior sequence as input and output one (or more) dense vector representing the user's interests, and an item encoder $g_{\theta}(y)\in \mathbb{R}^d$, that represents the items in the same vector space as the user encoder. We denote all the trainable parameters in the system as $\theta$, which includes the parameters in $f_{\theta}$ and $g_{\theta}$. With the learned encoders and the extracted item vectors, \ie, $\{g_{\theta}(y)\}_{y\in\mathcal{Y}}$, a k-nearest-neighbor search service, e.g., Faiss \cite{JDH17}, will be deployed for Top-N recommendation.
Specifically, at serving time, an arbitrary user behavior sequence $x$ will be transformed into a vectorial representation $f_{\theta}(x)$ and top $N$ items with the largest matching scores will be retrieved as Top-N candidates. Such matching scores are typically computed as inner product $\phi_\theta(x,y)=\langle f_{\theta}(x), g_{\theta}(y) \rangle$ or cosine similarity. In a nutshell, the learning procedure can be formulated as the following maximum likelihood estimation:
\begin{align}
	\arg\min_\theta \frac{1}{|\mathcal{D}|}\sum_{(x,y)\in \mathcal{D}} -\log p_\theta(y \mid x), \label{eq:sim}
    \quad \\
    \mathrm{where}\; p_\theta(y\mid x)=\frac{\exp{\phi_\theta(x,y)}}{\sum_{y'\in \mathcal{Y}}{\exp{\phi_\theta(x,y')}}}.
    \label{eq:mle}
\end{align}
In the matching phase, it can be infeasible to sum over all possible items $y'$ as in the denominator. Here we adopt the sample softmax objective \cite{Bengio_Senecal_2008,Jean_Cho_Memisevic_Bengio_2015}.
To demonstrate the effectiveness of the proposed CauseRec architecture, we utilize a naive framework as the baseline. Specifically, the \textit{item} encoder $g_{\theta}(y)$ is a plain lookup embedding matrix where the $n$th vector represents the item embedding with item id $n$. The \textit{user} encoder $f_{\theta}(x)$ aggregates the embeddings of historically interacted items using global average pooling and then transforms the aggregated embedding into the same embedding space as item embeddings using multi-layer perceptrons (MLP):
\begin{align}
	f_{\theta}(x) = \mathrm{MLP}(\frac{1}{t-1} \sum_{i=1}^{t-1}g_{\theta}(y_i)). \label{eq:interestencoder}
\end{align}
\subsection{The CauseRec Architecture}

%--------------------------------fig---------------------
\begin{figure*}[t] \begin{center}
    \includegraphics[width=0.9\textwidth]{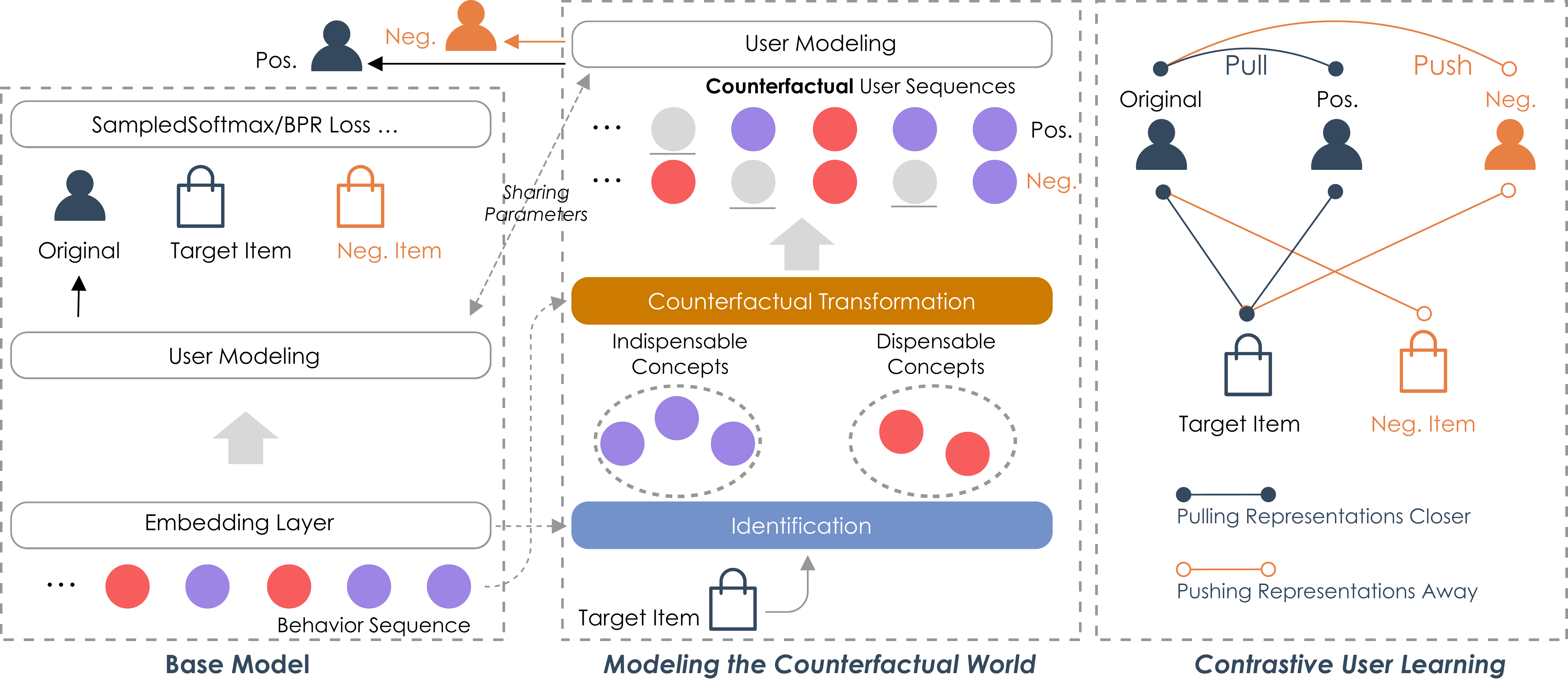}
    \caption{
    	Schematic of the proposed CauseRec-Item framework. 
	}
\vspace{-0.2cm}
\label{fig:schema}
\end{center} \end{figure*} 
%--------------------------------fig end---------------------

In this section, we give an brief illustration on the intuition and overall schema/pipeline of the CauseRec architecture, which is depicted in Figure \ref{fig:schema}, and introduce the building blocks in detail.

\subsubsection{Overall Schema} \label{sec:overallSchema}

The essence of CauseRec is to answer the retrospect question, 'what the user representation would be if we intervene on the observed behavior sequence?' The counterfactual transformation in CauseRec relates to the 'intervention on the observed behavior sequence.' For answering 'what the user representation would be,' we introduce an important inductive bias that makes the intervention work as expected. Specifically, we first identify indispensable/dispensable concepts in the historical behavior sequence. An indispensable concept indicates a subset of one behavior sequence that can jointly represent a meaningful aspect of the user's interest. A dispensable concept indicates a noisy subset that is less meaningful/important in representing an aspect of interest. We introduce the details in Section \ref{sec:identify}.

Given the identified concepts, a representative counterfactual transformation is designed to build out-of-distribution counterfactual user sequences. Here comes the inductive bias, \ie, counterfactual sequences constructed by replacing the dispensable concepts in the original user sequence should still have similar semantics to the original one. Here \textit{semantics} refer to the characteristics of user interests/preferences. Therefore, replacing indispensable concepts in the original user sequence should incur a preference deviation from the resulted user representation to the original user representation. We denote these resulted user representations as counterfactually \textit{negative} user representations. We note that such negatives are hard negatives where \textit{hard} refers to that other dispensable concepts stay the same as the original user sequence and \textit{negatives} means that the semantics of the user sequence should be different. In contrast, replacing dispensable concepts in the original user sequence should incur no preference change in representations. We denote these resulted user representations as counterfactually \textit{positive} user representations.
Different contrastive learning objectives are proposed to learn accurate and robust user representations that are less sensitive to the (attack on the) dispensable/noisy concepts and that trust more on the indispensable concepts that better represent the user's interest. Details can be found in Section \ref{sec:contrast}.

\subsubsection{Identification of Indispensable/Dispensable concepts} \label{sec:identify}

\begin{sloppypar}
To identify indispensable/dispensable concepts, we propose to first extract concept proposals and compute the proposal scores.

\end{sloppypar}
	\vpara{Item-level Concepts.} Inspired by instance discrimination \cite{He_Fan_Wu_Xie_Girshick_2020}, a straightforward while workable solution is to treat each item in the behavior sequence as an individual concept since each item has its unique fine-grained characteristics. In this way, we obtain the concept sequence $\mathbf{C} = \mathbf{X} \in \mathbb{R}^{t \times d}$, where $\mathbf{X} = g_{\theta}(x_{u,t+1})$ denotes the vectorial representations of the behavior sequence. In essence, concept scores indicate to what extent these concepts are important to represent the user's interest. Since there is no groundtruth for one user's real interest, we use the target item $y_{t+1}$ as the indicator:
\begin{align}
	p^{item}_i = \phi_{\theta} \left( \mathbf{c}_i , \mathbf{y} \right),
\end{align}
where $\mathbf{c}_i$ indicates the representation of $i$th concept in $\mathbf{C}$, and $\mathbf{y}$ indicates the representation of the target item. $\phi_{\theta}$ is the similarity function, and we empirically use dot product for its effectiveness in the experiment. $p^{item}_i$ is thus the score for the $i$th concept.

\vpara{Interest-level Concepts.}However, such a solution may incur redundancy in concepts since some items may share similar semantics, and might deteriorate the capability of modeling higher-order relationships between items. To this end, besides the item-level concepts, we introduce interest-level concepts by leveraging the attention mechanism \cite{Vaswani_Shazeer_Parmar_Uszkoreit_Jones_Gomez_Kaiser_Polosukhin_2017} to extract interest-level concepts. Formally, $\mathbf{X} \in \mathbb{R}^{t \times d}$, we obtain the following attention matrix:
\begin{align}
	\mathbf{A}=\operatorname{softmax}\left(\mathbf{W}_{2} \tanh \left(\mathbf{W}_{1} \mathbf{X}^{\top}\right)\right)^{\top}, \label{eq:interest}
\end{align}
where $\mathbf{W}_{1} \in \mathbb{R}^{d_a \times d}$ and $\mathbf{W}_{2} \in \mathbb{R}^{K \times d_a}$ are trainable transformation matrices. K is thus the number of concepts that is pre-defined. $\mathbf{A}$ is of shape $\mathbb{R}^{t \times K}$.
We obtain the concept sequence as the following:
\begin{align}
	\mathbf{C} = \mathbf{A}^{\top}\mathbf{X}, \label{eq:interestconcept}
\end{align}
Since interest-level concepts and the target item are not naturally embedded in the same space, we compute the concept score by the weighted sum of item-level scores:
\begin{align}
	P^{interest} = \mathbf{A}^{\top} \phi_{\theta}(\mathbf{X}, \mathbf{y}).
\end{align}
For both item-level concepts and interest-level concepts, we treat the top half concepts with the highest scores as indispensable concepts and the remaining half concepts as dispensable concepts. This strategy is mainly designed to prevent the number of indispensable or dispensable concepts from being too small. We leave finding more effective solutions as future works as illustrated in Section \ref{sec:conclusion}.

\subsubsection{Counterfactual Transformation} \label{sec:countertrans}

The proposed counterfactual transformation aims to construct out-of-distribution user sequences by transforming the original user sequence for one user. We here use \textbf{\textit{user sequence}} to generally denote the concept sequence, which can be either the commonly known item-level concept sequence, \ie, the original user behavior sequence, or the interest-level concept sequence. Based on the inductive bias described in Section \ref{sec:overallSchema}, we propose to replace the identified indispensable/dispensable concepts at the rate of $r_{rep}$ to construct counterfactually negative/positive user sequences, respectively. We note that directly dropping indispensable/dispensable concepts also seems feasible, but replacement has the advantage of not affecting overall sequence length and the relative positions of remaining concepts. We maintain a first-in-first-out queue as a concept memory for each level and use dequeued concepts as substitutes. We enqueue the concepts extracted from the current mini-batch. We denote the user sequence with indispensable/dispensable concepts being replaced as counterfactually negative/positive user sequence.

\begin{sloppypar}
	\subsubsection{User Encoders} \label{sec:userencoder} We note that item-level concept representations can be trained along with the item embedding matrix in most recommendation frameworks. Therefore, the above item-level concept identification and counterfactual transformation processes can be performed without any modification on the user encoder in the original baseline model, \ie, a model-agnostic and non-intrusive design. We denote the architecture solely considering item-level concepts as CauseRec-Item. CauseRec-Item obtains counterfactually positive/negative user representations $\{\mathbf{x}^{+,m}\}_{m=1,\dots,M}$/$\{\mathbf{x}^{-,n}\}_{n=1,\dots,N}$ from counterfactual item-level concept sequences using the original user encoder $f_\theta$.
\end{sloppypar}

\begin{sloppypar}
	We denote the architecture solely considering interest-level concepts as CauseRec-Interest. Different from CauseRec-Item, interest-level concepts are constructed with learnable parameters, \ie, $\mathbf{W}_1$ and $\mathbf{W}_2$ in Equation \ref{eq:interest}. Therefore, CauseRec-Interest is an intrusive design, and the inputs to the user encoder should be the interest-level concept sequence rather than the behavior sequence at the item-level. We note that there are no further modifications, and the architecture of the user encoder can stay the same as in the original baseline model. CauseRec-Interest obtains counterfactually positive/negative user representations from counterfactual interest-level concept sequence using the original user encoder $f_\theta$.
	
	We denote the architecture that considers counterfactual transformation on both the item-level concept sequence and the interest-level concept sequence as CauseRec-H(ierarchical). CauseRec-H is also an intrusive design with interest-level concepts as the inputs of the user encoder. Different from CauseRec-Interest, CauseRec-H further considers counterfactual transformations performed on item-level concepts. The counterfactually transformed item-level sequence will be forwarded to construct interest-level concept sequence using Equation \ref{eq:interest}-\ref{eq:interestconcept}. We note that counterfactual transformations will not be performed on these two levels simultaneously, which might introduce unnecessary noises. In other words, each counterfactual user representation is constructed with transformation on sequence solely from one level.
\end{sloppypar}

\subsubsection{Learning Objectives.} \label{sec:contrast} Besides the original recommendation loss $\mathcal{L}_{matching}$ described near Equation \ref{eq:sim}, we propose several contrastive learning objectives that are especially designed for learning accurate and robust user representations.

\vpara{Contrast between Counterfactual and Observation.} As discussed in Section \ref{sec:overallSchema}, a \textbf{\textit{robust}} user representation should be less sensitive to (possible attack on) dispensable concepts. Therefore, the user representations learned from counterfactual sequences with indispensable concepts transformed should be intuitively pushed away from the original user representation. Similar in spirit, an \textbf{\textit{accurate}} representation should trust more on indispensable concepts. Therefore, user representations learned from counterfactual sequences with dispensable concepts transformed should be intuitively pulled closer to the original user representation. Under these intuitions, we derive inspiration from the recent success of contrastive learning in CV \cite{He_Fan_Wu_Xie_Girshick_2020,Wu_Xiong_Yu_Lin_2018} and NLP \cite{Clark_Luong_Le_Manning_2020}, we use triplet margin loss to measure the relative similarity between samples:
\begin{align}
	\mathcal{L}_{co}  &= \sum_{m=1}^{M} \sum_{n=1}^{N} \max \left\{d\left(\mathbf{x}^{q}, \mathbf{x}^{+,m}\right)-d\left(\mathbf{x}^{q}, \mathbf{x}^{-,n}\right)+\Delta_{\mathrm{co}}, 0\right\},
\end{align}
where $\mathbf{x}^{q}$ denotes the original user representation. We set the distance function $d$ as the L2 distance since user representations generated by the same user encoder are in the same embedding space. We empirically set the margin $\Delta_{\mathrm{co}}=1$.

\vpara{Contrast between Interest and Items.} The above objective considers the user representation side solely, and we further capitalize on the target item $y_t$, which also enhances the user representation learning. Formally, given the L2-normalized representation of the target item $\mathbf{\tilde y}$ and user representation $\mathbf{\tilde x}$, we have:
\begin{align}
%	\mathbf{\tilde x} &= \mathbf{x} / \left\| \mathbf{x} \right\|_{2} \\
%&1- \mathbf{\tilde x}^{q} \cdot \mathbf{\tilde y}_t +
	\mathcal{L}_{ii} =  \sum_{m=1}^M 1- \mathbf{\tilde x}^{+,m} \cdot \mathbf{\tilde y} + \sum_{n=1}^N \max \left( 0, \mathbf{\tilde x}^{-,n} \cdot \mathbf{\tilde y} -\Delta_{\mathrm{ii}} \right), \label{eq:ii}
\end{align}
%where $\mathbf{\tilde x}^{q}$ is the L2-normalized original user representation. 
This objective also has the advantage of preventing the user encoder from learning trivial representations for counterfactual user sequences. We set the margin $\Delta_{\mathrm{ii}}=0.5$ in the experiment.
Finally, the loss for training the whole framework can be written as:
\begin{align}
	\mathcal{L}_{cause} = \mathcal{L}_{matching} + \lambda_1 \mathcal{L}_{co} + \lambda_2 \mathcal{L}_{ii}.
\end{align}
During testing/serving, only the backbone model that generates the user representation is needed. The identification of indispensable/dispensable concepts and the counterfactual transformation processes are disregarded. Noteworthy, the computation of proposal scores which depends on the target item does not belong to the backbone model and is not required during testing.

\section{Experiments}

We conduct experiments on real-world public datasets and mainly aim to answer the following three research questions:

\begin{itemize}[leftmargin=*]
	\item \textbf{RQ1}: How does CauseRec perform compared to the base model and various SOTA sequential recommenders?
	\item \textbf{RQ2}: How do the proposed building blocks and different hyper-parameter settings affect CauseRec?
	\item \textbf{RQ3}: How do user representations benefit from modeling the counterfactual world and contrastive representation learning?
\end{itemize}

%--------------------------------table---------------------
\begin{table}[!t]
\caption{Statistics of the Datasets.}
\centering
% \textbf{Model Predictive Control}
\begin{tabular}{l cc cc}
\toprule
Dataset & \#Users & \#Items & \#Interactions & \#Density  \\
% \cline{2-5} 
    \midrule
    Amazon Books          & 459, 133 & 313, 966 & 8, 898, 041 & 0.00063 \\ 
    Yelp   & 31, 668 & 38, 048 & 1, 561, 406 & 0.00130 \\
    Gowalla   & 52, 643 & 91, 599 & 2, 984, 108 & 0.00084 \\
    \bottomrule
\end{tabular}%
    \label{tab:staData}
\vspace{-0.2cm}
\end{table}
%
%--------------------------------table end---------------------

%--------------------------------table---------------------
\begin{table*}[h]
\centering
    \caption{Comparison results of three CauseRec architectures with SOTA sequential recommenders designed for the matching phase. CauseItem/CauseIn/CauseH stand for CauseRec -Item/-Interest/-Hierarchical, respectively. The symbol $*$ indicates the improvements over the strongest baseline (underlined) are statistically significant ($p < 0.05$) with one-sample t-tests.}
{\setlength{\tabcolsep}{0.63em}\begin{tabular}{ll cccccccc cccc}
\toprule
  Datasets  & Metric & POP & Y-DNN  & GRU4Rec &  MIND & ComiSA & ComiDR  & CauseItem & CauseIn & CauseH & Improv.   \\
    \midrule
   
  \multirow{6}{*}{Books} 
   &  R@20  &  0.0137  &  0.0457   &  0.0406    & 0.0486  &  \underline{0.0549}  &  0.0531  &  0.0582  &  0.0593  &  \textbf{0.0623*}  & 13.5\%  \\
   &  R@50  &  0.0240  &  0.0731   &  0.0650    & 0.0764  &  \underline{0.0847}  &   0.0811   &  0.1001  &   0.0993  &  \textbf{0.1018*}   & 20.2\%  \\
   &  NDCG@20  &  0.0226  &  0.0767   &  0.0680   & 0.0793   &  0.0899 & \underline{0.0918}    &  0.0985  &   0.1006  &  \textbf{0.1051*}  &  14.5\% \\
   &  NDCG@50  &  0.0394  &  0.1208   &   0.1037   & 0.1223  &  \underline{0.1356}  & 0.1352   &  0.1628  & 0.1619   &  \textbf{0.1655*}  & 22.1\%  \\
   &  HR@20  & 0.0302   &  0.1029   &   0.0894   &  0.1062 &  0.1140  &\underline{0.1201}    & 0.1280   &    0.1303  &  \textbf{0.1370*}   & 14.1\%  \\
   &  HR@50  & 0.0523   &  0.1589   &   0.1370   &  0.1610 &  0.1720  & \underline{0.1758}   &  0.2078  &  0.2062   &  \textbf{0.2113*}   & 20.2\%  \\
    \midrule

    \multirow{6}{*}{Yelp} 
   &  R@20  	&	0.0016 &  0.0506   &  0.0454  &   0.044    &  \underline{0.0534} 	 &  0.0472 	  & 0.0570   & 0.0580  &  \textbf{  0.0591*  }  &  10.7\% \\
   &  R@50  	&   0.003 &   0.1048  &  0.0937  &   0.0943    &   \underline{0.1101}    &   0.0935     &  0.1163  &  0.1175	 &  \textbf{  0.1182*  } &  7.36\% \\
   &  NDCG@20  	&   0.0065 &  0.1582  &  0.1447 &   0.1414    &   \underline{0.1728} 	&  0.1453    &  0.1812  & 0.1806  	 &  \textbf{  0.1830*  }  & 5.90\%  \\
   &  NDCG@50  	&   0.0129 &  0.2887  &   0.2673 &  0.2699    &   \underline{0.3025} 	 &  0.2612   &  0.3179  &  0.3180	 &  \textbf{  0.3210*  }  & 6.12\%  \\
   &  HR@20  	&   0.0152 &  0.3015   &  0.2826  &   0.2681    &  \underline{0.3249} 	 &  0.2775   &  0.3416  &  0.3391 	 &  \textbf{  0.3426*  }  & 5.45\%  \\
   &  HR@50  	&   0.0268 &  0.5131   &  0.4853  &  0.4866    &   \underline{0.5324} 	 &  0.4629 	  &  0.5583  & 0.5576  &  \textbf{  0.5605*  }  & 5.28\%  \\
    \midrule
    \multirow{6}{*}{Gowalla}

   &  R@20  	&  	0.0028  & 	0.1127   &  0.1273	 &    0.1218	   &   \underline{0.1277}  &   0.1153     & 0.1315   &   0.1355  & \textbf{0.1359*}   &  6.42\% \\
   &  R@50  	&  	0.0054  &  	0.1926  & 0.2043 	 &  	 0.2049   & \underline{0.2072}    &  0.1831      & 0.2238   &   0.2204    &  \textbf{0.2251*}    & 8.64\%  \\
   &  NDCG@20 	 &  0.0073	  & 	0.2378   &  \underline{0.2803}	&    0.2565  	     & 0.2736    &   0.2534    &  0.2747  &    0.2825  &  \textbf{0.2842*}  & 1.39\%  \\
   &  NDCG@50  	&   0.0135	 &  	0.3638       & 0.4002     & 	      0.3888    &  \underline{0.4019}   &   0.3621   &  0.4123  &  0.4113  & \textbf{0.4221*}   & 5.03\%  \\
   &  HR@20 	 &  	0.0104  &  	0.3443  & 0.3814 	&     0.3627   	    &  \underline{0.3838}   &   0.3429     &  0.3918  &   0.3995   &   \textbf{0.4042*}  &  5.32\% \\
   &  HR@50 	 &  	0.0224  & 	0.5010   &	 0.5251 &   0.5301       &  0.5288   &   \underline{0.5355}    &  0.5596  &    0.5553     &  \textbf{0.5697*}  & 6.39\%  \\
    \bottomrule
\end{tabular}}
    \label{tab:comparison}
\vspace{-0.1cm}
\end{table*}
%
%--------------------------------table end---------------------

\subsection{Experimental Setup}

To demonstrate the generalization capability on learning users' representations of the proposed CauseRec architecture, we employ an evaluation framework \cite{Liang_Krishnan_Hoffman_Jebara_2018,Ma_Zhou_Cui_Yang_Zhu_2019,Cen_Zhang_Zou_Zhou_Yang_Tang_2020} where models should confront unseen user behavior sequences. Specifically, the users of each dataset are split into training/validation/test subset by the proportion of $8:1:1$. For training sequential recommenders, we incorporate a commonly used setting by viewing each item in the behavior sequence as a potential target item and using behaviors that happen before the target item to generate the user's representation, as defined in Section \ref{sec:problemformulation}. For evaluation, only users in the validation/test set are considered, and we choose to generate users' representations on the first 80\% behaviors, which are unseen during training. Such a framework can help justify whether models can learn accurate and robust user representations that can generalize well. We mainly focus on the \textit{matching} phase of recommendation and accordingly choose the datasets, comparison methods, and evaluation metrics.

\vpara{Datasets} We consider three challenging recommendation datasets, of which the statistics are shown in Table \ref{tab:staData}.
\begin{itemize}[leftmargin=*]
\item \textbf{Amazon Books.} We take Books category from the product review datasets provided by \cite{McAuley_Targett_Shi_Hengel_2015}, for evaluation. For each user, we keep at most 20 behaviors that are chronologically ordered.
\item \textbf{Yelp2018.} Yelp challenge (2018 edition) releases the review data for small businesses (\eg, restaurants). We view these businesses as items and use a 10-core setting \cite{Wang_He_Wang_Feng_Chua_2019,He_Deng_Wang_Li_Zhang_Wang_2020} where each item/user has at least ten interactions. 
\item \textbf{Gowalla.} A widely used check-in dataset \cite{Liang_Charlin_McInerney_Blei_2016} from the Gowalla platform. Similarly, we use the 10-core setting \cite{He_McAuley_2016}.
\end{itemize}

\vpara{Comparison Methods} We mainly consider sequential recommenders for comparison since models are required to confront unseen behaviors for each user. Therefore, factorization-based and graph-based methods are not considered. The compared state-of-the-art models are listed as the following:
\begin{itemize}[leftmargin=*]
	\item \textbf{POP.} A naive baseline that always recommends items with the most number of interactions.
	\item \textbf{YouTube DNN \cite{Covington_Adams_Sargin_2016}.} A successful industrial recommender that generates one user's representation by pooling the embeddings of historically interacted items.
	\item \textbf{GRU4Rec \cite{Hidasi_Karatzoglou_Baltrunas_Tikk_2016}.} An early attempt to introduce recurrent neural networks into recommendation. 
	\item \textbf{MIND \cite{Li_Liu_Wu_Xu_Zhao_Huang_Kang_Chen_Li_Lee_2019}.} The first framework that extracts multiple interest vectors for one user based on the capsule network.
	\item \textbf{ComiRec-DR \cite{Cen_Zhang_Zou_Zhou_Yang_Tang_2020}.} A recently proposed SOTA framework following MIND to extract diverse interests using dynamic routing and incorporate a controllable aggregation module to balance recommendation diversity and accuracy.
	\item \textbf{ComiRec-SA \cite{Cen_Zhang_Zou_Zhou_Yang_Tang_2020}.} ComiRec-SA differs from ComiRec-DR by using self-attention to model interests.
\end{itemize}

\vpara{Evaluation Metrics} We employ three broadly used numerical criteria for the matching phase, \ie, \textit{Recall}, \textit{Normalized Discounted Cumulative Gain} (NDCG), and \textit{Hit Rate}. 
We report metrics computed on the top 20/50 recommended candidates. Higher values indicate better performance for all metrics.

\vpara{Implementation Details} We use Adam \cite{Kingma_Ba_2015} for optimization with learning rate of 0.003/0.005 for Books/Yelp and Gowalla, $\beta_1= 0.9$, $\beta_2 = 0.99$, $\epsilon=1 \times 10^{-8}$, weight decay of $1 \times 1e-5$. We train CauseRec-Item for (maximum) 10 epochs and CauseRec-Interest/CauseRec-H for (maximum) 30 epochs with mini-batch size 1024. All models are with embedding size 64. We set hyper-parameters $\lambda_1 = \lambda_2 = 1$ and do not tune them with bells and whistles. As illustrated in Section \ref{sec:base}, the item encoder is a plain embedding lookup matrix, and the user encoder is a three-layer perceptron with hidden size 256. We set $N=8$, $M=1$, $r_{rep}=0.5$ for CauseRec-Item/-Interest and $N=16$ $M=2$ for CauseRec-H to accommodate transformation on two levels, as illustrated in Section \ref{sec:userencoder}. We set $K=20$ for CauseRec-Interest/-H.

\subsection{Performance Analysis (RQ1)}

The comparison results of CauseRec with SOTA sequential recommenders are listed in Table \ref{tab:comparison}. We report three architectures of CauseRec including CauseRec-Item (CauseItem), CauseRec-Interest (CauseIn), and CauseRec-Hierarchical (CauseH), as described in Section \ref{sec:userencoder}. In a nutshell, we observe a clear improvement of these architectures over various comparison methods and across three different metrics. Notably, CauseRec-H improves the previous SOTA ComiRec-SA/DR by +.0299 (relatively 22.1\%) concerning NDCG@50 on the Amazon Books dataset and +.0179 (relatively 8.64\%) concerning Recall@20 on the Gowalla dataset. Among the comparison methods, ComiRec mostly yields the best performance by modeling multiple interests for a given user. However, only modeling the noisy historical behaviors might result in diverse but noisy interests that may not accurately represent users, finally leading to inferior results. GRU4Rec achieves comparably good results with ComiRec on the Gowalla dataset. GRU4Rec can effectively model the sequential dependency between items in the behavior sequence. However, it might be more likely to suffer from the noises due to the strict step-by-step encoding process. In contrast, CauseRec architectures confront the noises within users' behaviors by pushing the user representation away from counterfactually negative user representations and pulling it closer to counterfactually positive user representations. Besides, these results demonstrate the generalization capability of CauseRec on confronting out-of-distribution user sequences by modeling the counterfactual world.

Among three CauseRec architectures, CauseRec-Item is a model-agnostic and non-intrusive design, which means it can be applied to any other sequential recommender without any modification on the original user encoder, and solely functions in the training stage without sacrificing inference efficiency. CauseRec-Interest constructs interest-level concepts by grouping items that may belong to a certain interest (\eg, \texttt{chocolate} and \texttt{cake} belong to \texttt{sweets}) into one holistic concept. Compared to CauseRec-Item, CauseRec-Interest has the advantage of reducing concept redundancy and modeling higher-order relationships between items, and thus improving CauseRec-Item. To combine the merits of CauseRec-Interest and CauseRec-Item, CauseRec-Hierarchical considers both interest-level and item-level concepts in counterfactual transformation. CauseRec-H achieves the best results, which shows that counterfactual transformation on item-level concepts still yields some unique advantages, such as modeling fine-grained preferences. For example, people might not generally like all \texttt{sweets} and prefer \texttt{cake} to \texttt{chocolate}.

\subsubsection{Ablation Studies.} We are interested in the CauseRec-Item architecture due to its strengths of being: easy to implement (model-agnostic), efficient in serving (non-intrusive), and effective. To obtain a better understanding of different building blocks in CauseRec-Item, we consider surgically removing some components and construct the following architectures. The results on Yelp and Gowalla datasets are shown in Table \ref{tab:ablation}.

%--------------------------------table---------------------
\begin{table*}[h]
\centering
    \caption{Ablation studies by constructing different architectures. We progressively ablate key components in CauseRec-Item, which is a model-agnostic and non-intrusive design.}
\begin{tabular}{l ccc ccc ccc ccc}
\toprule
&\multicolumn{6}{c}{Yelp } & \multicolumn{6}{c}{Gowalla }  \\
\midrule
Model & \multicolumn{3}{c}{Metrics@20} & %
    \multicolumn{3}{c}{Metrics@50} & \multicolumn{3}{c}{Metrics@20} & %
    \multicolumn{3}{c}{Metrics@50} \\
% \cline{2-5} 
\iffalse

0.3851	0.5398	0.1256	0.209	0.2693	0.3945

\fi

\cmidrule(lr){2-4}\cmidrule(lr){5-7}\cmidrule(lr){8-10}\cmidrule(lr){11-13}
    & Recall & NDCG & Hit Rate & Recall & NDCG & Hit Rate & Recall & NDCG & Hit Rate & Recall & NDCG & Hit Rate \\
    \midrule
    CauseRec-Item     & 0.0570	 & 0.1812	 & 0.3416	 & 0.1163  & 	0.3179 	 & 0.5583   & 0.1315	 & 0.2747	 & 0.3918	 & 0.2238  & 	0.4123 	 & 0.5596	\\ 
    \midrule % 
    w.o. $\mathcal{L}_{co}$    & 	0.0563 & 0.1794		& 0.3350	  & 	0.1169 & 0.3174 & 0.5507	 & 	0.1286 & 0.2719		& 0.3865	  & 	0.2153 & 0.4055 & 0.5522	\\ 
    w.o. $\mathcal{L}_{ii}$       & 0.0518		 & 0.1679  & 0.3113	 & 0.1067	 & 0.2983	 & 0.5267 & 0.1256		 & 0.2668  & 0.3758	 & 0.2110	 & 0.3968	 & 0.5445	\\ 
    \midrule % 0.1286	0.2719	0.3865	0.2153	0.4055	0.5522
   	$Pos$ Only       & 	0.0518 & 0.1662 & 	0.3101	 & 0.1047	 & 	0.2911 & 	0.5115  	& 	0.1256	 & 0.2693	 & 	0.3851	  & 	0.2090		 & 		0.3945 & 	0.5398\\ 
    $Neg$ Only & 	0.0494 & 0.1633 & 	0.3044  & 	0.1024	 & 0.2909	& 	0.5137 	  & 	0.1246	 & 	0.2608 & 	0.3754		 & 	0.2053	 & 	0.3925	 & 	0.5368  	\\ 
    \midrule
    
    Base Model     & 	0.0444 & 0.1444 & 	0.2722	 & 0.0934	 & 	0.2650 & 	0.4692   & 	0.1208 & 0.2569 & 	0.3670	 & 0.2000	 & 	0.3811 & 	0.5218\\ 

    % 
    % base model relative improvement %
    \bottomrule
\end{tabular}
    \label{tab:ablation}
\vspace{-0.2cm}
\end{table*}
%--------------------------------table end---------------------

\vpara{w.o. (without) $\mathcal{L}_{co}$.} This means we do not consider the contrast between the counterfactual and the observation. The performance drop compared to CauseRec-Item indicates that pushing counterfactually negative user representation away and pulling counterfactually positive user representation closer can potentially help the learned observational user representation to trust more on indispensable items (accurate) and to be immune from dispensable items (robust).

\vpara{w.o. $\mathcal{L}_{ii}$.} $\mathcal{L}_{ii}$ is defined in Equation \ref{eq:ii}. Removing $\mathcal{L}_{ii}$ means we do not consider the contrast between counterfactual user representations and positive/negative target items. According to Table \ref{tab:ablation}, we observe a clear performance drop compared to CauseRec-Item. Furthermore, eliminating $\mathcal{L}_{ii}$ yields poorer performance than eliminating $\mathcal{L}_{co}$. We attribute this to that, $\mathcal{L}_{ii}$ prevents the user encoder from yielding trivial representations for counterfactual user sequences (item-level and interest-level) by contrasting counterfactual user representations with target items representations. In other words, $\mathcal{L}_{co}$ with possibly trivial counterfactually user representations (without $\mathcal{L}_{ii}$) might hurt the effectiveness $\mathcal{L}_{co}$.

\vpara{$Pos$ Only.} CauseRec-Item with only counterfactual transformations on dispensable items and counterfactually positive user representation is denoted as $Pos$. This architecture disregards $\mathcal{L}_{co}$ and term $\sum_{n=1}^N \max \left( 0, \mathbf{\tilde x}^{-,n} \cdot \mathbf{\tilde y}_t -\Delta_{\mathrm{margin}} \right)$ in $\mathcal{L}_{ii}$. Not surprisingly, $Pos$ Only architecture achieves inferior results compared with w.o. $\mathcal{L}_{co}$ architecture, demonstrating the merits of counterfactually negative user representations. Still, this architecture improves the base model, demonstrating the effectiveness of contrasting counterfactual user representations with target items for recommendation.

\vpara{$Neg$ Only.} CauseRec-Item with only counterfactual transformations on indispensable items is denoted as $Neg$. We observe similar results to the $Pos$ Only architecture. This again verifies the effectiveness of the contrastive user representation learning framework by modeling the counterfactual distribution. Compared to $Pos$ Only architecture, $Neg$ Only architecture achieves inferior results. This phenomenon might indicate that making user representation trust more on indispensable concepts can be potentially more important than eliminating the effect of indispensable concepts (\ie, noisy items in CauseRec-Item) on user representation learning. This is reasonable in the sense that the former process might potentially make the user in the embedding space away from all other items, including the dispensable items that are replaced during counterfactual transformation.

\vpara{Base Model} A naive matching baseline described in Section \ref{sec:base}. All other Architectures yield improvement over the Base Model.

\subsubsection{Analysis on the number of counterfactual user representations.}

We ablate the number of counterfactually positive/negative user representations, \ie, M and N, as defined in Section \ref{sec:userencoder}. As shown in Table \ref{tab:numposneg}, we can observe that 1) increasing the number (from $N=M=1$ to $N=M=8$) not necessarily leads to better performance. 2) particularly increasing the number of counterfactually negative user representations (from $N=M=1$ to $N=8, M=1$) can be useful. This result is reasonable in the sense that such representations can be interpreted as hard negatives since each of the corresponding counterfactual user sequences contains dispensable items staying the same as the original user sequence, and hard negatives are known to be helpful. 3) particularly increasing the number of counterfactually positive user representations (from $N=M=1$ to $N=1, M=8$) may introduce noises since each of them contains some randomly sampled items that can be falsely interpreted as "positive". Therefore one counterfactually positive user representation can be enough for learning accurate user embeddings.

\subsubsection{Analysis on the Replace Ratio.}

We are interested in how the replacement ratio, \ie, $r_{rep}$, in counterfactual transformation of CauseRec-Item affects the model performance. Table \ref{tab:replacerate} shows the results by varying $r_{rep}$. The best result is achieved with $r_{rep}=0.4/0.5$ for the Yelp dataset and $r_{rep}=0.5$ for the Gowalla dataset. Either too small or too large $r_{rep}$ will lead to sub-optimal results. We attribute this phenomenon to that small $r_{rep}$ will affect the capability of counterfactual learning and large $r_{rep}$ will introduce more noises brought by randomly sampled items for replacement. $r_{rep}$ makes a tradeoff between these two impacts.

\subsubsection{Analysis on the number of interest-level concepts.}

Here we take an analysis on the number of interest-level concepts particularly for CauseRec-Interest architecture. Specifically, we ablate $K$ as described in Equation \ref{eq:interest}. As shown in Table \ref{tab:analysisK}, we observe a performance improvement with $K$ increasing (\eg, $4\rightarrow10$, $10\rightarrow20$). Each interest-level concept can be more coarse-grained when $K$ becomes smaller and more find-grained when $K$ becomes larger. It can be hard to classify a coarse-grained concept as indispensable or dispensable. It may lead to also coarse-grained or even inaccurate transformations afterward, and thus hurting the performance. Too large $K$ (\eg, 30) may bring redundancies and noises (more random concepts introduced with a fixed replace ratio) to the framework and eventually leads to inferior results.

%--------------------------------table---------------------
\begin{table}[!t]
\centering
    \caption{Performance analysis on the number of counterfactually positive/negative user representations in CauseRec-Item, denoted in $M$/$N$.}
{\setlength{\tabcolsep}{0.45em}
\begin{tabular}{l ccc ccc}
\toprule
&\multicolumn{3}{c}{ Yelp } &\multicolumn{3}{c}{ Gowalla } \\
\midrule
Model
   & R@50 & N@50  & H@50  & R@50 & N@50  & H@50  \\
    \midrule
    $N=1, M=1$    &  0.111    &  0.310    &  0.541   &  0.219    &  \textbf{0.413}    &   0.559         \\
    $N=4, M=4$    &  0.113    &  0.308    &   0.542   &  0.210    &  0.394    &   0.541      \\
    $N=8, M=8$    &  0.106    &  0.298    &   0.524   &  0.204    &  0.381    &   0.521  \\
    $N=8, M=1$    &  \textbf{0.116}    &  \textbf{0.318}    &  \textbf{0.558}   &  \textbf{0.224}    &  0.412    &   \textbf{0.560}    \\
    $N=1, M=8$    &  0.096    & 0.273     &   0.483    &  0.185    & 0.361     &   0.498   \\
    \bottomrule
\end{tabular}}%

    \label{tab:numposneg}
\vspace{-0.2cm}
\end{table}
%--------------------------------table end---------------------

%--------------------------------table---------------------
\begin{table}[!t]
\centering
    \caption{Performance analysis on the replace rate $r_{rep}$ in counterfactual transformation for CauseRec-Item.}
\begin{tabular}{l ccc ccc}
\toprule
&\multicolumn{3}{c}{ Yelp } &\multicolumn{3}{c}{ Gowalla } \\
\midrule
Model
   & R@50 & N@50  & H@50  & R@50 & N@50  & H@50  \\
    \midrule
    $r_{rep} = 0.2$    &   0.115   &  \textbf{0.318}    &   0.552    &   0.209   &  0.389    &   0.531        \\
    $r_{rep} = 0.4$    &   \textbf{0.117}   &  \textbf{0.318}    &    0.556   &   0.209   &  0.401    &   0.543        \\
    $r_{rep} = 0.5$    &   0.116   &  \textbf{0.318}    &    \textbf{0.558}   &   \textbf{0.224}   &  \textbf{0.412}    &   \textbf{0.560}        \\
    $r_{rep} = 0.6$    &   0.113   &  0.307    &   0.537    &   0.210   &  0.397    &   0.537        \\
    $r_{rep} = 0.8$    &   0.106   &  0.296    &   0.520    &   0.211   &  0.399    &   0.544        \\
    \bottomrule
\end{tabular}%

    \label{tab:replacerate}
\vspace{-0.2cm}
\end{table}
%--------------------------------table end---------------------

%--------------------------------table---------------------
\begin{table}[!t]
\centering
    \caption{Analysis on the number of constructed interest concepts $K$ for CauseRec-Interest.}
\vspace{-0.2cm}
{\setlength{\tabcolsep}{0.65em}\begin{tabular}{l ccc ccc}
\toprule
&\multicolumn{3}{c}{ Yelp } &\multicolumn{3}{c}{ Gowalla } \\
\midrule
Model
   & R@50 & N@50  & H@50  & R@50 & N@50  & H@50  \\
    \midrule
    $K=4$    &   0.100   &  0.28    &   0.501    &   0.206   &  0.390    &   0.531        \\
    $K=10$    &   0.111   &   0.305   &  0.536     &   0.215   &  0.403    &   0.546        \\
    $K=20$    &   \textbf{0.118}   &   \textbf{0.318}   &   \textbf{0.558}    &   \textbf{0.220}   &  \textbf{0.411}    &   \textbf{0.555}         \\
    $K=30$    &   0.117   &   0.311   &   0.547    &   0.219   &  0.406    &   0.547        \\
    \bottomrule
\end{tabular}}%

    \label{tab:analysisK}
\vspace{-0.4cm}
\end{table}
%--------------------------------table end---------------------

%--------------------------------fig-------------------------
\begin{figure}[!t] \begin{center}
\begin{subfigure}{.235\textwidth}
	\includegraphics[width=1\linewidth]{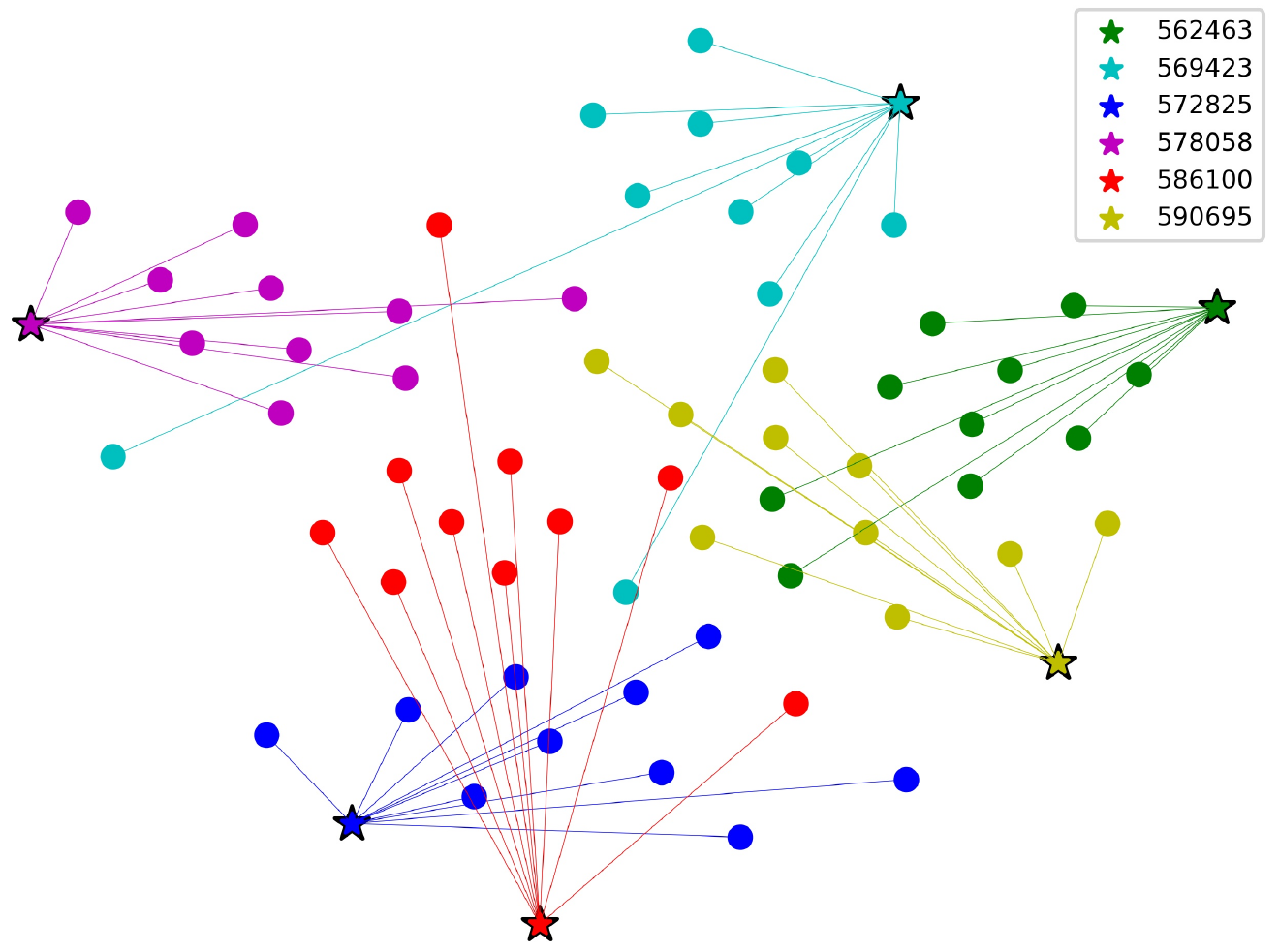}
    \caption{
    \footnotesize{Base Model}
    	}
%\vspace{-0.4cm}
\label{fig:tsnebase}
\end{subfigure}
\begin{subfigure}{.235\textwidth}
	\includegraphics[width=1\linewidth]{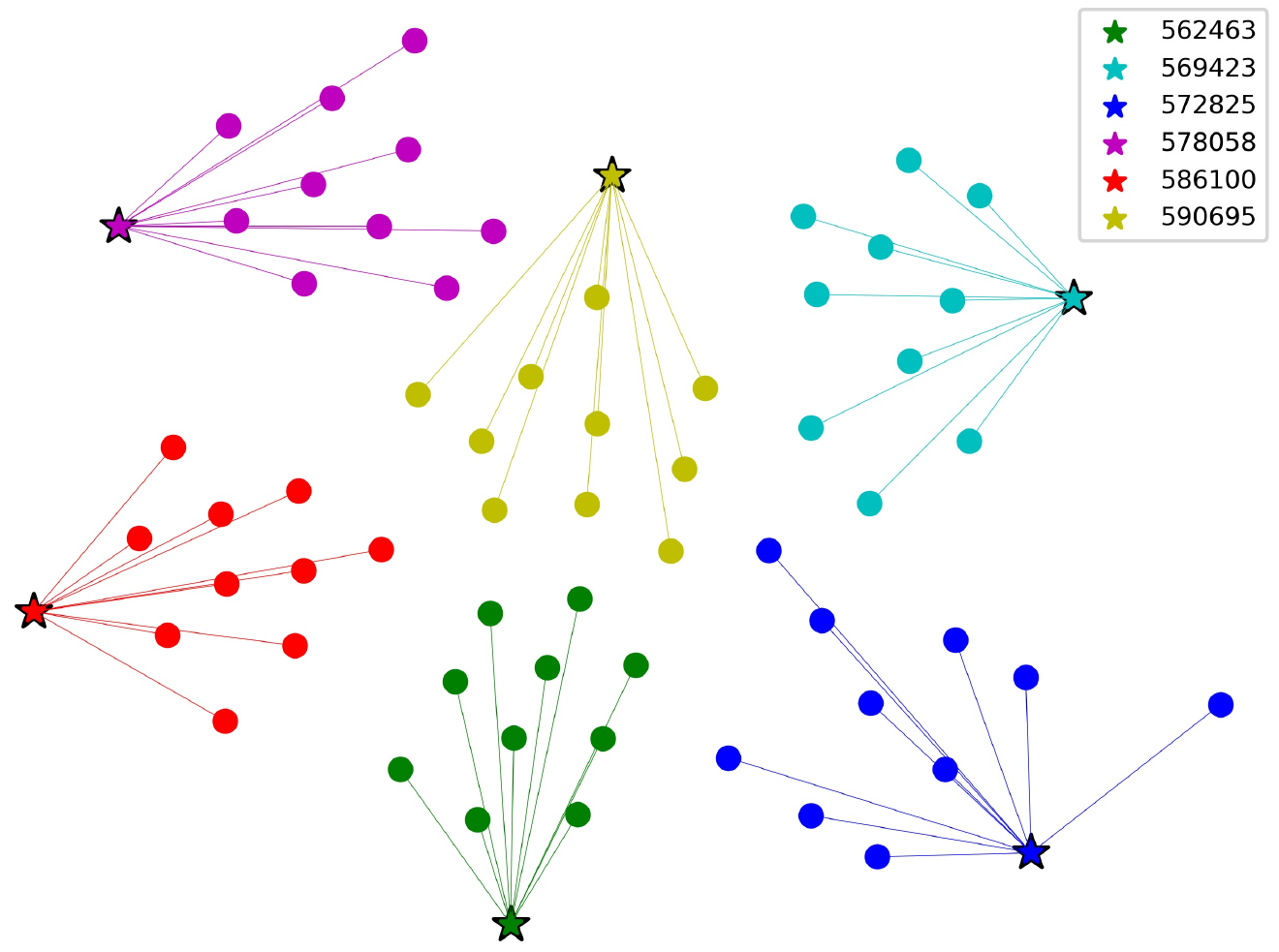}
    \caption{
   \footnotesize{CauseRec}
    	}
%\vspace{-0.4cm}
\label{fig:tsneCauseRec}
\end{subfigure}
    \caption{
    Visualization of randomly sampled users (shown as stars) with their interacted items (shown as points of the same color) from the Amazon Books dataset. We perform the t-SNE transformation on the representations learned by the base model (left) and CauseRec (right).
    	}
    \label{fig:tsne}
%\vspace{-0.2cm}
\end{center} \end{figure}
%--------------------------------fig end--------------------	

\subsection{Case Study (RQ3)}

To understand how the learned user representations benefit from the CauseRec framework, we plot randomly selected six users from the Amazon books dataset. We also plot ten corresponding items sampled from the test set for each user. Specifically, we perform t-SNE transformation on the user/item representations learned by the base model (as shown in Figure \ref{fig:tsnebase}) and CauseRec-Item (as shown in Figure \ref{fig:tsneCauseRec}). The connectivities of users and test items in the embedding space can help reflect whether the model learns accurate and robust user representations. From Figure \ref{fig:tsneCauseRec}, we observe that users with their corresponding test items easily form clusters and show small intra-cluster distances and large inter-cluster distances. By jointly comparing the same users (\eg, 590695, and 586100) in Figures \ref{fig:tsnebase} and \ref{fig:tsneCauseRec}, we can see that CauseRec-Item helps the user encoder learn representations that are closer to their corresponding test items. These results qualitatively demonstrate the effectiveness of CauseRec on learning accurate and robust user representations.

\begin{figure}[!t] \begin{center}
    \includegraphics[width=0.8\columnwidth]{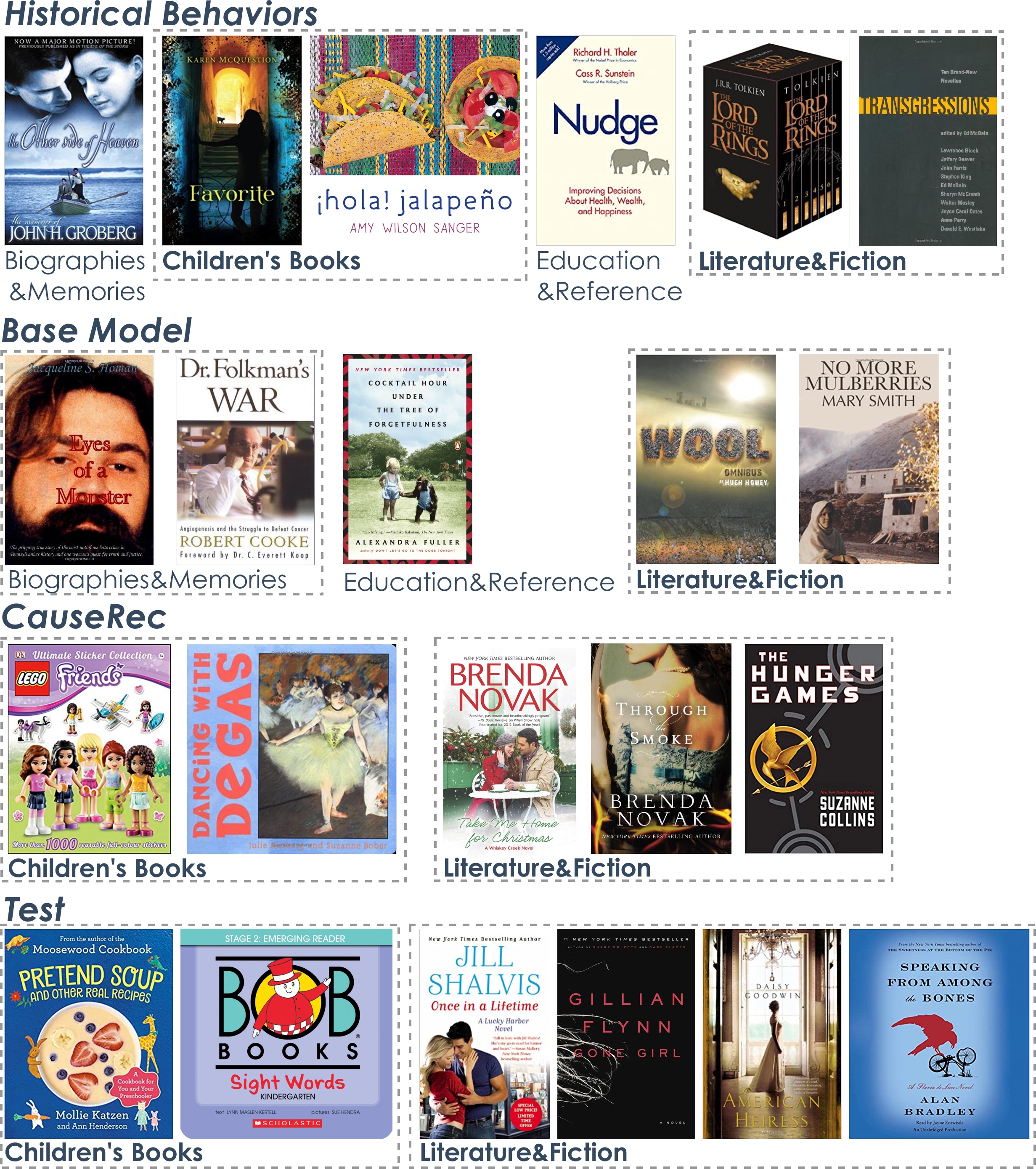}
    \caption{
    	Case study by visualizing a real-world sample from the Amazon Books testing set. We mainly show the books' covers and categories for clarity.
	}
\label{fig:case}
%\vspace{-0.2cm}
\end{center} \end{figure}

We also present a recommendation result from the Amazon Books test datasets in Figure \ref{fig:case}. We list the historical behaviors, the top five books recommended by the base model and CauseRec-Item, and books interacted by the corresponding user in the test set. We mainly visualize the books' covers and categories for better clarity. We note that the side information is generally not considered in training matching models (both the base model and CauseRec). As shown in Figure \ref{fig:case}, we observe that CauseRec yields more consistent recommendation results to the books in the test set. Supposing historical behaviors consist of noisy ones, and behaviors in the test accurately reflect users' interest for the current state, CauseRec successfully captures users' interests, \ie, Children's Books, and Literature\&Fictions. In contrast, the base model is more likely to be affected by noisy behaviors that appear only a few times, such as the Biographies\&Memories, and Education\&Reference. These results further demonstrate that CauseRec can learn accurate and robust user representations that are less distracted by noisy behaviors.

%--------------------------------fig---------------------
%--------------------------------fig end---------------------

\section{Conclusion and Future Work} \label{sec:conclusion}

\begin{sloppypar}
	In this work, we propose to model the counterfactual data distribution to confront the sparsity and noise nature of observed user interactions in recommender systems. The proposed CauseRec conditionally sample counterfactually positive and negative user sequences with transformations on the dispensable/indispensable concepts. We propose multiple structures (-item, -interest, -hierarchical) to confront both fine-grained item-level concepts and abstract interest-level concepts. Several contrastive objectives are devised to contrast the counterfactual with the observational to learn accurate and robust user representations. Among several proposed architectures, CauseRec-Item has the advantage of being non-intrusive, \ie, solely functioning at training while not affecting serving efficiency. With a naive matching baseline, CauseRec achieves a considerable improvement over it and SOTA sequential matching recommenders. Extensive experiments help to justify the strengths of CauseRec as being both simple in design and effective in performance.
\end{sloppypar}

This work can be viewed as an initiative to exploit the joint power of constative learning and counterfactual thinking for recommendation. We believe that such a simple and effective idea can be inspirational to future developments, especially in model-agnostic and non-intrusive designs. CauseRec-Item is compatible with various user encoders within most existing sequential recommenders. We choose a naive baseline to better demonstrate the effectiveness of this work, and we plan to explore its strengths in more models. Another future direction is to whether more effective solutions of identifying indispensable/dispensable concepts exist, including both the computation of concept scores and the determination of indispensable or dispensable for each concept based on the scores. Lastly, we will explore the strengths of CauseRec for the ranking phase of recommendation. Counterfactual transformations designed with various auxiliary features and complex model architectures will open up new research possibilities.

\section{ACKNOWLEDGMENTS}

\begin{sloppypar}
The work is supported by the National Key R\&D Program of China (No. 2020YFC0832500), NSFC (61625107, 61836002, 62072397), Zhejiang Natural Science Foundation (LR19F020006), and Fundamental Research Funds for the Central Universities (2020QNA5024).
\end{sloppypar}

\newpage
\newpage
\balance
\bibliographystyle{ACM-Reference-Format}
\bibliography{sections/9.citations}

\end{document}